# Data-Intensive Workload Consolidation on Hadoop Distributed File System

Reza Moraveji
School of Information Technologies
The University of Sydney and National ICT Australia (NICTA)
Sydney, Australia
reza.moraveji@sydney.edu.au

Javid Taheri
School of Information Technologies
The University of Sydney
Sydney, Australia
javid.taheri@sydney.edu.au

MohammadReza HosseinyFarahabady
School of Information Technologies
The University of Sydney and National ICT Australia (NICTA)
Sydney, Australia
mreza@it.usyd.edu.au

Nikzad Babaii Rizvandi
School of Information Technologies
The University of Sydney and National ICT Australia (NICTA)
Sydney, Australia
nikzad@it.usyd.edu.au

Albert Y. Zomaya
School of Information Technologies
The University of Sydney
Sydney, Australia
albert.zomaya@sydney.edu.au

*Abstract*— Workload consolidation, sharing physical resources among multiple workloads, is a promising technique to save cost and energy in cluster computing systems. This paper highlights a few challenges of workload consolidation for Hadoop as one of the current state-of-the-art data-intensive cluster computing system. Through a systematic step-by-step procedure, we investigate challenges for efficient server consolidation in Hadoop environments. To this end, we first investigate the inter-relationship between last level cache (LLC) contention and throughput degradation for consolidated workloads on a single physical server employing Hadoop distributed file system (HDFS). We then investigate the general case of consolidation on multiple physical servers so that their throughput never falls below a desired/predefined utilization level. We use our empirical results to model consolidation as a classic two-dimensional bin packing problem and then design a computationally efficient greedy algorithm to achieve minimum throughput degradation on multiple servers. Results are very promising and show that our greedy approach is able to achieve near optimal solution in all experimented cases.

## I. INTRODUCTION

Recently, data-intensive cluster computing systems have increasingly become important to perform a wide range of applications including –but not limited to– machine learning, data mining, and image/text processing [1]. MapReduce [2] is among the most well-known cluster computing frameworks directly benefited from consolidation technologies to perform its heavy data-intensive applications. Hadoop [3], an open-source version of Google's MapReduce, is a reliable and cost-effective framework for data-intensive distributed computing applications. This framework is built on a large-scale cluster storage managed by Hadoop distributed file system (HDFS) [4]; HDFS is designed for storing very large files on clusters of commodity hardware where the chance of node failure is high [1].

Data centers benefited from consolidation through various ways. Firstly, consolidation is aligned with recent trends in data center management which aims to reduce resource cost and improve resource utilization [5]. Secondly, it is one of the most important techniques to conserve energy in cloud computing environments [13] where physical servers are aimed to maintain well utilized without compromising throughput of concurrent workloads more than a threshold. Poor workload consolidation, on the other hand, may lead to high resource contention, and consequently unbalanced distribution of workloads among nodes; i.e., some computational nodes may attain significantly worse throughput and utilization than others [6, 22]. For example, job latency on Facebook's Hadoop clusters started to become unacceptably high when a wrong mixture of production daily, ad hoc, and real-time jobs were consolidated on them [1]. Poor consolidation can also paralyze an entire Hadoop cluster and put production jobs at risk [8]. Performance unpredictability for run-times of MapReduce jobs on EC2 cluster is another example of inefficient workload consolidation [7].

From workload consolidation point of view, Shared resources such as last level processor cache (LLC) in multicore physical servers have always showed unique challenges to seamless adoption of servers in distributed computing environments [9]. While sharing such resources through increasing resource utilization is generally beneficial, lack of control over concurrent workloads can significantly lead to unacceptable loss of throughput and unpredictable response time of individual workload [10].

The objective of our study in this paper is to experimentally investigate how to load shared resources of a cluster of servers with data-intensive applications so that their throughput degradation never falls below a threshold. To achieve this, firstly, we investigate the throughput of a single workload on a single physical server. We show that system parameters such as LLC, disk cache, and system file cache are the main bottlenecks to maintain high throughput; also, throughput varies according to two application-specific parameters: file size and request size. Secondly, we examine throughput of multiple workloads when combined on a single physical server. Results of these experiments are then used to model the effect of LLC contention, disk bandwidth, and processor execution time on throughput. Finally, we use the results from our second step and generalize our problem for workload consolidation of multiple workloads on multiple physical servers. In this step, we also formulate the general workload consolidation as a two-dimensional bin packing problem and design a greedy algorithm to solve it.

The paper is organized as follows. Section II surveys related works. Section III analyses throughput of a single workload on a single server. Section IV studies the throughput of multiple co-run workloads on a single physical server and provides mathematical models for that. Section V introduces two constraints to guarantee achieving minimum makespan for coallocated workloads. In section VI, we study the general case of multiple workloads on multiple servers. In section VII, we propose a greedy algorithm for server consolidation. Section VIII explained our experimental evaluation followed by section IX that concludes our work.

## II. RELATED WORKS

Workload consolidation has been a thoroughly studied topic for cluster computing systems, especially to investigate the tradeoff between workload consolidation and throughput degradation. These studies usually consider different types of workloads –e.g., CPU-intensive and data-intensive–, different performance goals –e.g., throughput, response time, and power–, and different frameworks –e.g., MapReduce, Dryad [12] etc. Therefore, we only summarize works that were closely related to the topics of interest in this paper.

A close work to our study is the task scheduler proposed in [5] in which a scheduler designed to predict the performance of concurrent MapReduce workloads and adjust their resources so that job response times are minimized. Delay scheduling [8] addresses the job latency problem on Hadoop clusters at Facebook and focuses on studying the tradeoff between fairness in scheduler and data locality in Hadoop applications. Quincy [11], a platform-specific scheduler implemented on Dryad distributed execution engine, is a fair-share scheduler also addressing the same problem. Authors of [7] address the problem of performance unpredictability and variance in EC2 cloud for MapReduce applications and discover that unpredictability is greatly related to poor workload consolidation.

Apart from the aforementioned works, there are also other works where consolidation is used to optimize power and energy. Energy-aware workload consolidation in [13] is an example attempts to conserve energy for disk-/CPU-intensive applications in cloud computing environments; their approach, however, lacks accurate workload characterization. In [14], a novel runtime framework is proposed to dynamically consolidate instances from different workloads into a single GPU workload; they also propose GPU performance and power models for effective workload consolidation on GPUs. Joulemeter [15] that is initially designed as a tool for power usage measurement of virtual machines aims to consolidate multiple workloads on fewer servers for improving resource utilization and power costs.

Analyzing the effect of last level processor cache (LLC) on workload consolidation –another topic of interest we investigate in this work– is also covered by several studies. For example, authors of [16] study the behavior of consolidated workloads particularly on sharing caches across a variety of configurations. In [10] authors also study shared resource monitoring to understand resource usage and ways to improve overall throughput as well as quality of service of a data center. A mathematical model has also been proposed to predict the effect of cache contention on the performance of consolidated workloads [17].

After close examination of all these works, we noticed several shortcomings and decided to cover them in this article; thus, we can highlight our contribution in this work through the following items. The first difference of our work with previously reported studies mainly lies in the way we characterize data-intensive applications with two main parameters: file size and request size; such characterization is inspired by well-known filesystem benchmarking tools, namely Iometer [18], IOzone [19], TestDFSIO [20], and Bonnie++ [21]. Our second contribution is related to the Hadoop distributed file system that has been never properly covered in previous studies –to the best of our knowledge. We believe this is the first work that thoroughly analyses inter-relationship between workload consolidation, throughput degradation, and LLC contention for data-intensive applications employing HDFS. Our third contribution is to propose mathematical models for different aspects of this study based on imperial results from TestDFSIO [20].

## III. SINGLE WORKLOAD ON SINGLE SERVER

In this section, we measure the throughput of a single workload on a single physical server. Here, we show that throughput is a function of file size (FS) and file operation request size (RS) of the workload. RS is the amount of data that workload reads/writes from/to a file in a single file operation. Our experimental results show that increasing FS beyond LLC size noticeably degrades the throughput of workload.

### A. Workload Characterization

We conduct a series of experiments on two physical servers to capture the effect of FS and RS on throughput. The experiments are based on the intuition that data-intensive workloads can be characterized by FS and RS [18-21]. As expected, the throughput curves for all servers follow the same pattern by varying FS and RS for both read and write operation.

TABLE I. EXPERIMENTAL SETUP

| | | **Experimental Setup** | | | | |
|---|---|---|---|---|---|---|
| | | *processor* | *last level cache* | *memory* | *System File Cache* | *Disk cache* |
| Physical Servers | M1 | Core(TM) i7 CPU @2.00 GHz | 6MB | 8GB | 980MB | 12MB |
| | M2 | Core(TM)2 Duo @ 3.00 GHz | 6MB | 3GB | 455MB | 8MB |

### B. Experimental Setup

Table I shows the experimental setup for these two servers ($M_1$ and $M_2$). On all physical servers, system file cache and file buffering are always activated in operating systems; thus, workloads always interact with system file cache rather than system disk. Through enabling this feature write-back cache always delays flushing file data until triggered by cache manager –usually at predefined time intervals. Note that although system file cache is a feature of native filesystem not that of HDFS, it still can significantly impact the performance of writing/reading data to/from

HDFS. Also, disk cache –embedded memory in hard drive– is enabled, therefore, systems do not wait for any device to access the correct location on the disk to write the data; here, the disk controller rather sends an acknowledge to operating system and saves significant amount of time needed for actual writing on the disk. Therefore, both system file cache and disk cache can significantly increase workload throughput as they (1) act as a read-ahead buffer so that the data can be read-ahead for future requests, and (2) act as write-back cache that delays actual writings.

As stated earlier, we use HDFS as the default filesystem for our experiments; Hadoop is particularly designed for storing very large-sized files where large files are split into block-sized chunks (64MB by default) to be independently stored in the system. Each workload –i.e., a map task– in turn works on these block-sized chunks. In our experiments, to work with non-defualt block-sized chunks, we change the filesystem installation parameters.

### C. Experimental Results on One Single Server

Figure 1 and 2 plot the throughput of data-intensive workloads against FS (block-sized chunk) and RS for read and write operations on both $M_1$ and $M_2$ from Table I. It is worth noting that these figures show the FS of a Hadoop *task* (usually in order of 64MB) and not the FS of a Hadoop *job* that is usually in order of Terabytes. Both figures show how throughput is affected by FS and RS of each experiment.

For each RS, these figures show two/three throughput levels for read/write operations in each plate. From left to right: (1) the first/highest throughput levels are related to small FSs that can easily fit into LLC of the servers (6MB for both servers), (2) the second/intermediate throughput levels start when FS becomes greater than LLC however less than the summation of system file cache and disk cache; and, (3) the third/lowest level –only for write operation– starts when FS exceeds such summation where the actual disk I/O speed can also be observed. For example, in Figures (1)b and (2)b, the third throughput level starts around (980MB+12MB) and (455MB+8MB) that are the summation of system file cache and disk cache for M1 and M2, respectively.

These figures also show that throughput is always improved by increasing size of RS. In details, there are four components contributing to affect total access time to a disk –either read or write–, they are: controller access time, seek time, rotational latency, and data read/write time. In real systems, the portion related to the actual read/write is negligible compared to the other components which are overhead. Reading/writing 1MB of data with RS=1KB takes much more time than that of RS=512KB because overhead happens 1000 times for the former and 2 times for the latter case. Therefore, accessing disks with large RSs are always much more efficient than acceding disks for small ones. We also like to stress that based on our extensive experiments we noticed that LLC, system file cache, and disk cache parameters from physical servers and FS and RS parameters from workloads greatly affect throughput of a system. Therefore, they must all be carefully designed/selected to achieve high-throughput workload consolidation schemes.

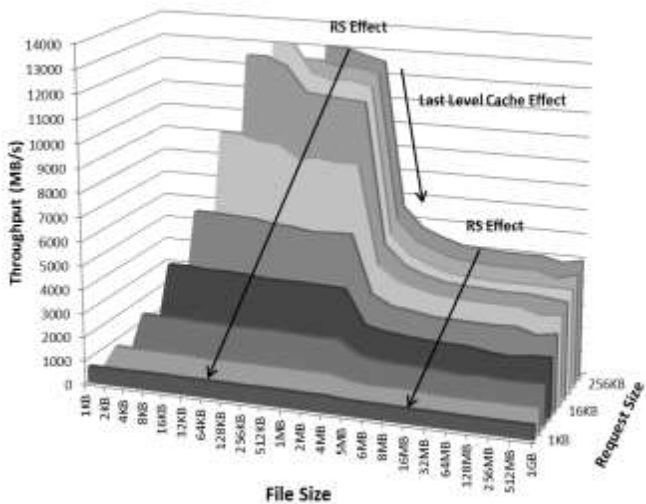
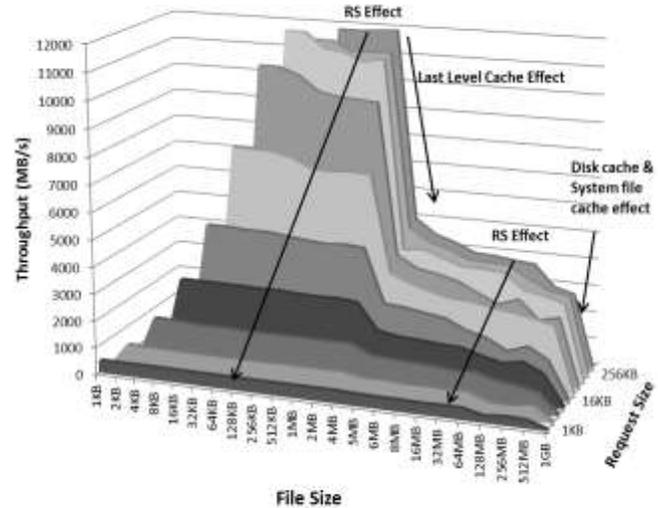

(a)            (b)
Figure 1. Single workload on a single server (a) read and (b) write operations on $M_1$.

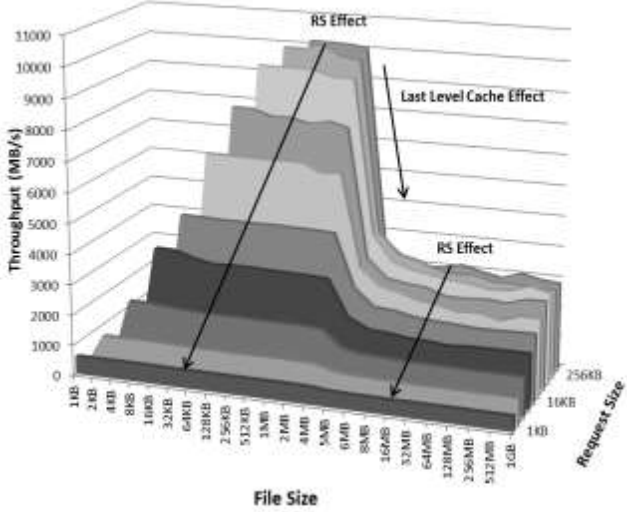 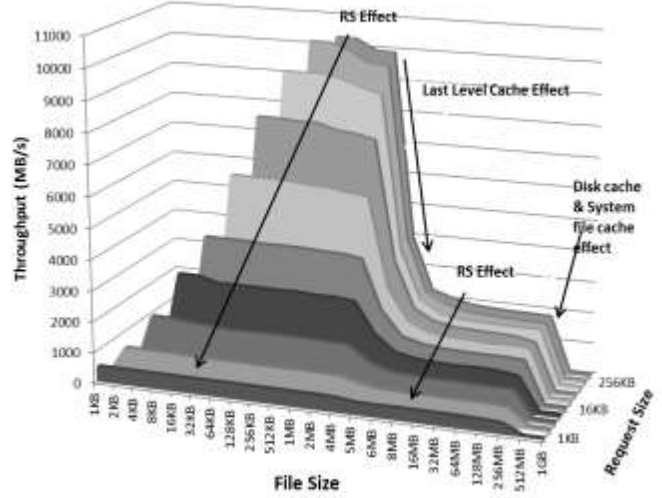

(a)                      (b)
Figure 2. Single workload on a single server (a) read and (b) write operations on $M_2$.

## IV. MULTIPLE WORKLOADS ON SINGLE SERVER

Upon our experiments for a single workload, we extend our experiments by measuring throughput of multiple workloads on a single physical server. Here, we consider different RSs, FSs, and number of concurrent workloads ($N$) to measure throughput. To present our results, we replace the "request size" axis of Figures 1 and 2 with the number of concurrent workloads to produce Figures 3 and 4 for this case. Because similar trend of throughput degradation were observed for different RSs, we only reflect result of RS=64KB and 256KB on $M_1$ server in this article. We also like define "saturation point" to refer to conditions where throughput degradation becomes greater than 50%; i.e., conditions in which the execution time of a workload is at least doubled. More discussion will be made in the following sections to highlight the importance of saturation points in achieving minimum makespans for consolidated workloads.

### A. Last Level Cache and Consolidation

Figure (3 and 4) shows that each plate gets a moderate slope till a particular FS (point) where throughput is sharply dropped. Such drop-off point in each plate is greatly related to condition where different workloads start to evict each others' cached data from their shared LLC. We will use the term "throughput degradation point" (TDP) to refer to such sharp degradation points for the rest of this paper. One of our aims in this work is to find/calculate TDPs for any combination of workloads and relate them to FSs and RSs.

Our observations showed that TDPs always occur when total amount of competing data for access LLC exceeds its capacity. In fact, the total amount of competing data to access LLC is a function of FSs and RSs of concurrent workloads. For instance, one of the TDPs in Figure 4(a) occurs when $N$=4, RS=256KB, and FS=1280KB; for this particular point, the total competing data to access LLC is around 4×(1280KB+256KB)=6MB: the exact cache size of $M_1$. This figure also shows that increasing any of these values (N,RS, or FS) always results in greater degradations of throughput as expected through our experimental model. Based on that we observed that TDP always occurs when:

$$CacheSize = \sum_{i=1}^{N}(RS_i + FS_i) \quad (1)$$

where $FS_i$ and $RS_i$ are the file and request size of workload $i$, respectively. Dotted points in Figure (3 and 4)a show calculated points of TDP from Eqn. (1) for each plate and graphically confirm our hypothesis for predicting sharp degradation points for concurrent workloads.

Our further observation also shows that Eqn. (1) always holds only when FS are smaller than LLC; otherwise, it will not compete to access LLC. Hence, if FS of a workload becomes greater than LLC, then it will not compete with others to access LLC. As a result, it should not be considered in calculating TDP. Based on such observations, we replace Eqn. (1) by the following:

$$CacheSize = \sum_{i=1}^{N}(RS_i) + \sum_{i=1 \& i \in CS}^{N}(FS_i)$$
$$CS = \{i \mid FS_i \leq CacheSize\} \quad (2)$$

to predict when TDP will occur.

## B. Mutual Throughput Degradation and Consolidation

Because LLC is not the only resource shared by consolidated workloads, other shared resources such as processor execution engine, system file cache, disk bandwidth, and disk cache can also cause throughput degradation if overloaded. The effect of these sources of throughput degradation can also be seen in Figures (3 and 4)a along the depth axis. The degradation is mostly related to competition of different workloads to access shared disk bandwidth and processor execution time. In fact, observed throughput is linearly degraded by increasing $N$. Therefore, we propose another model to predict the degradation caused by a group of workloads on a single workload, $j$, as follows:

$$D_j = \sum_{i=1, i \neq j}^{N} D_{i,j} \quad (3)$$

where $D_{i,j}$ is the throughput degradation caused by workload $i$ on $j$. Validation of our hypothesis (model) is illustrated in Figure (3 and 4)b. In the figures, we compare the actual throughput degradations with the ones predicted by the model for two RSs. Indeed, the model predicts the degradations with reasonable accuracy.

As a basis of our model, we need to collect all $D_{i,j}$'s through running $(10 \times 23) \times (10 \times 23) = 52900$ individual experiments to capture all possible combination of our experimental setups; i.e., ten RSs (1KB-512KB) and 23 FSs (1KB-1GB) for each workload.

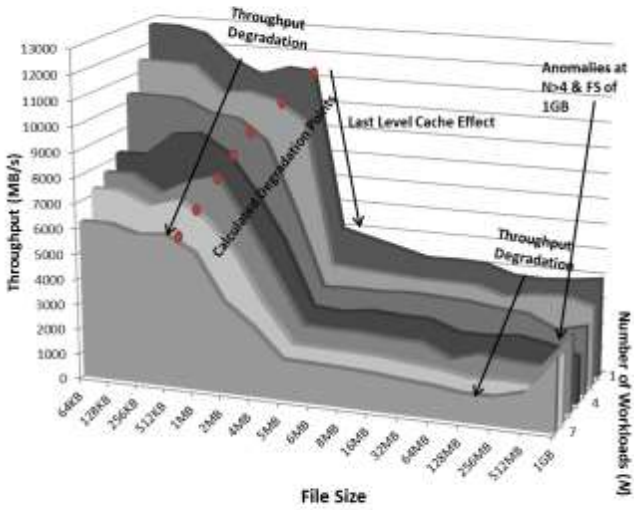

(a)

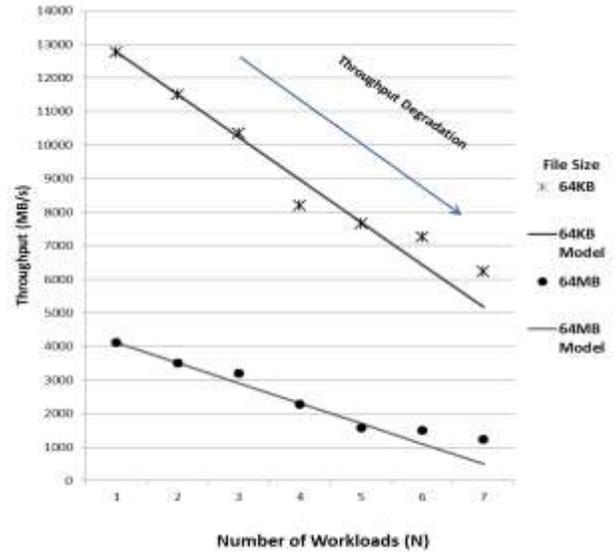

(b)

Figure 3. (a) Multiple workloads with 64KB request size on a single server ($M_1$) and (b) model validation for throughput degradation.

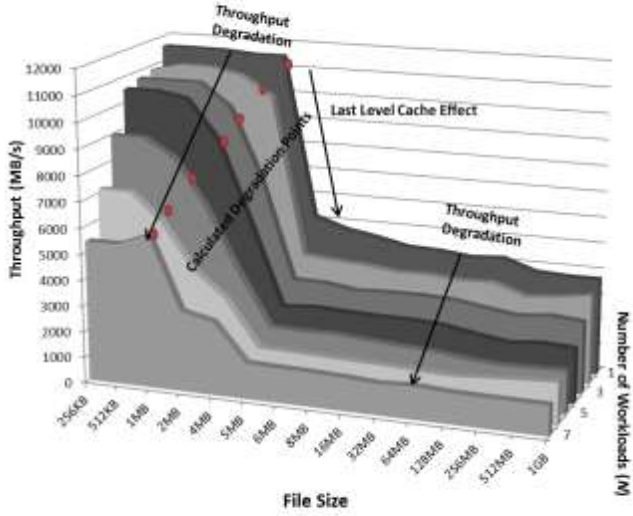
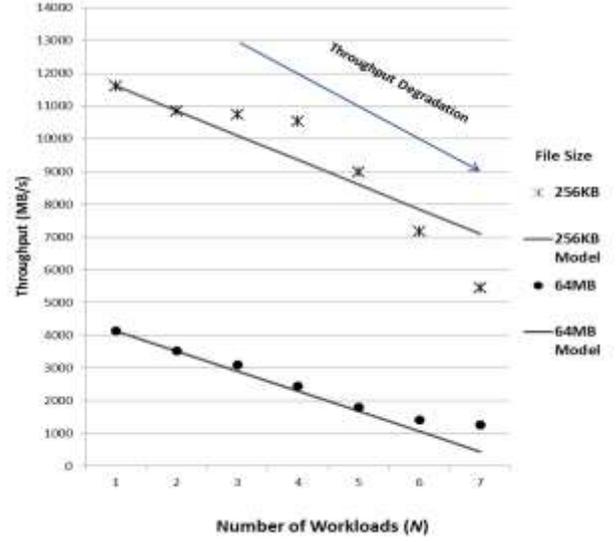

(a)                  (b)

Figure 4. (a) Multiple workloads with 256KB request size on a single server ($M_1$) and (b) model validation for throughput degradation.

## V. BOUNDS FOR CONSOLIDATION PROBLEM

Before formulating consolidation problem, we like to introduce two criteria to achieve desired throughput. The goal of the first criterion is to achieve minimum makespan by restricting the maximum number of consolidated workloads on a physical server. Our proposed criterion is to decide where a new workload must be run upon its arrival to the system; new workload is allocated to a server where throughput degradation of all its co-run workloads (including the new workload) becomes less than 50% after consolidation. If, no server is found to satisfy this criterion, then, the new workload will be queued until a server to satisfy this criterion is found –most probably upon completion of another workload. Here, we like to show how following this criterion always results in a lower makespan – compared with when it is not followed– for the consolidated workloads of each server. For better explanation, consider two scenarios depicted in Figure 5. In this figure, $AR_i$ represents the actual running time of workload i ($W_i$) when it is solely run on a physical server; $O_i$ represents the time overhead imposed on $W_i$ because of its coallocations to another workload. As can be seen, in the first co-run scenario, because $O_1<AR_1$ and $O_2<AR_2$, the makespan of colloating $W_1$ and $W_2$ is always less than $W_1 + W_2$ (sequential). The second co-run scenario shows the other possibility in which $O_3>AR_3$ and $O_4<AR_4$. As indicated in the figure, makespan of co-allocating $W_3$ and $W_4$ is ($W_3 + O_3$) that is greater than ($W_3 + W_4$). In other words, running $W_3$ and $W_4$ one after another is better than consolidating them!

The following equation mathematically formulates such situations and defines $D_i$, total degradation on workload $i$, variables to detect such situations.

$$D_i = \frac{O_i}{(AR_i + O_i)}$$

Using $D_i$'s, this criterion implied to consolidate only when:

$$\forall\, O_i, W_i \quad D_i = \frac{O_i}{(AR_i + O_i)} < 0.5$$

or

$$D_i = \sum_{j=1, j \neq i}^{N} D_{i,j} < 0.5 \quad (4)$$

For the scenarios in Figure 5, $D_1<0.5$ and $D_2<0.5$, while $D_3>0.5$ and $D_4<0.5$. Note that the aforementioned criterion can be useful only when consolidated workloads have identical run-times; i.e., $AR_1=AR_2=\ldots=AR_N$ for all workloads.

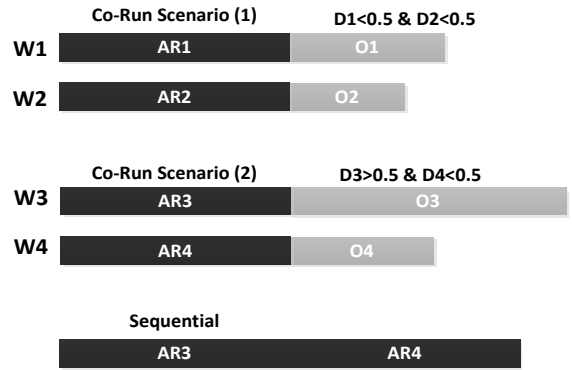

Figure 5. Constraint on the number of concurrent workloads.

The second criterion is to adjust the number of consolidated workloads so that their total cached requests do not exceed LLC's capacity. We already show in Figure (3 and 4)a that throughput degradation as a result of losing cache is noticeable. Here, we experimentally show, Figure 6(a and b), that this throughput degradation is always more than 50%. In the figures, curves on top refer to workloads that could efficiently access LLC; the ones at the bottom represent those that have lost such competition to constantly access LLC in their favor. The figures reveal that for RSs>8KB, throughput degradation is always more than 50%. Therefore, we design our second criterion to check/estimate throughput degradation before consolidation as: consolidate workloads only when

$$\sum_{i=1}^{N}(RS_i) + \sum_{i=1 \& i \in CS}^{N}(FS_i) \leq \alpha \times CacheSize \quad (5)$$
$$CS = \{i \mid FS_i \leq CacheSize\}$$

$\alpha$ manifests a threshold level in which a system cache can be overloaded. For example, $\alpha=1.2$ for cache size of 12MB means the cache can tolerate concurrent accesses of multiple workloads with total data of 14.4MB without significant throughput degradation. $\alpha$ can be empirically found through comparing the actual TDPs of a system versus its calculated ones. In our case, for example, in Figure (3 and 4)b actual TPDs are around 7.76MB, whereas the calculated TDPs are 6MB. Thus, for our system $\alpha$ should be about $7.76/6 \approx 1.3$.

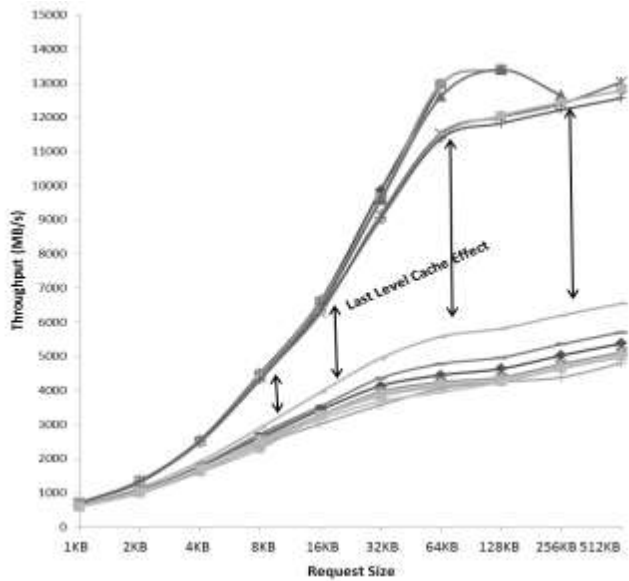
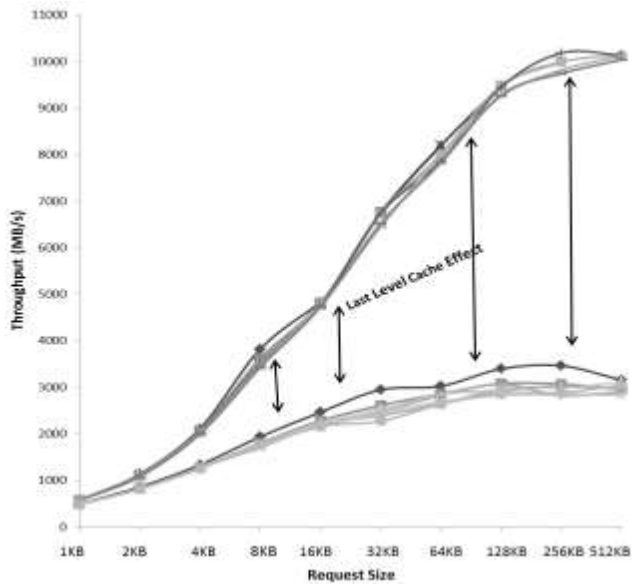

Figure 6. The effect of losing last level cache on throughput degradation

## VI. CONSOLIDATION PROBLEM: MULTIPLE WORKLOADS ON MULTIPLE SERVERS

In this section, we formulate consolidation problem as a two-dimensional bin packing problem –based on the direct observations from the previous sections– so that throughput of no workload in degraded for more than 50%. To this end, physical servers are modeled as 2D-bins to adjust the number of consolidated workloads on a physical server. The first dimension is inspired by our first criterion to check/estimate that throughput of individual workloads never falls below 50%; the second dimension is inspired by the second criterions to check/estimate that the total amount of competing data for LLC is always bounded by $\alpha.CacheSize$. Figure 7 shows a graphical representation of such dimensions. Here, each workload is defined as an object to be packed into one of the server bins; the first and second dimension for each workload is defined as its 'FS and RS' and 'mutual throughput degradation on other workloads', respectively. It is worth noting that our formulated problem is much harder than the original multi-dimensional bin packing problem as objects are independent in the original case, whereas mutually affecting each other on our formulation.

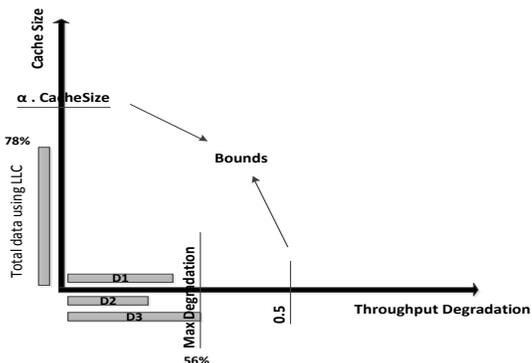

Figure 7. A typical physical server, bin, after allocation

## VII. CONSOLIDATION ALGORITHM

This section presents a greedy algorithm (Figure 8) to solve the consolidation problem formulated in the previous section. The aim of our greedy algorithm is to minimize throughput degradation of consolidated workloads when distributing them among a given number of physical servers. The greedy minimizes the sum of the average loads on both dimensions on all physical servers after allocation. The load on one dimension is the total amount of competing data for LLC and the load on the other dimension is the maximum of throughput degradations. To better explain the loads consider the three consolidated workloads in Figure 7 again. In this example, $D_3$ has the maximum throughput degradation of 56% which is representative of the load in second dimension and the total amount of competing data if 78% of $α×CacheSize$ denoting the load in the first dimension of the respective bin.

```
Greedy Algorithm
comments:
    m: number of physical servers
    S_i: physical server i
    D_{x,y}: throughput degradation of workload x on y
Input: W_j
0.  minimum = 100%
1.  for i := 1 to m do
        begin
2.          Assign W_j to S_i
3.          CacheInUse_i =
                (total amount of competing data on S_i)
                                /
                (α_i * CacheSize_i)
            Comment: maximum throughput degradation
                on S_i, Max(D_y) is calculated based on
                Previously collected D_{x,y}s on the server
            Calculate Max(D_y)
4.          Comment: check if the allocation does not
                violate 50% degradation rule
            If Max(D_y) > 50% or CacheInUse_i > 100% then
                Go to 1.
5.          Avg_i = Avg(CacheInUse_i, Max(D_y))
6.          If Avg_i < minimum then
                begin
                    minimum := Avg_i
                    S_i is the candidate of allocation
                end
        end
7.  Allocate W_j → S_i
```

Figure 8. The proposed greedy algorithm

To better explain our greedy algorithm in Figure 8, assume two physical servers, A and B, have loads according to Table II. For servers A/B, workloads occupy 30%/40% of $α×CacheSize$ with maximum throughput degradation of 40%/45%. Now assume a new workload, W, is arrived. If W is allocated to A/B, then, the total amount of data will be 35%/42% of $α×CacheSize$ with maximum throughput degradation of 45%/48%. In this case Avg(A before)+Avg(B after) will be (30+40)/2+(42+48)/2=80, whereas Avg(A before)+Avg(B after)=82.5. Therefore, W will be consolidated with the current load in B assigned to it. As can be seen, our greedy algorithm tries to consolidate workloads so that summation of all servers' degradation is minimized. It is also worth noting that such allocation greatly depends on the sequence of arriving workloads. However, finding the optimal solution through brute-force search can heavily overload schedulers and thus almost impossible to implement.

TABLE II. AN EXAMPLE FOR THE GREEDY ALGORITHM

| | Servers | %Total Data of Cache | %Maximum Degradation | Average |
|---|---|---|---|---|
| Initial States | A (before) | 30 | 40 | 35 |
| | B (before) | 40 | 45 | 42.5 |
| After Allocating | A (after) | 35 | 45 | 40 |
| | **B (after)** | 42 | 48 | 45 |
| | | Avg A (before) + Avg B (after) | Avg B (before) + Avg A (after) | |
| | | **80** | 82.5 | |

## VIII. EXPERIMENTAL EVALUATION

To evaluate the performance of the proposed greedy algorithm, we implement the algorithm on a cluster of four servers: $2xM_1s$ and $2xM_2s$ from Table I. Hadoop nodes are running Ubuntu 11.10 as detailed in section III/B. We also developed a light-weight program to monitor maximum throughput degradation on each server with sampling time interval of 1s. Our workload generator is inspired by Iometer [18], IOzone [19], TestDFSIO [20], and Bonnie++ [21]; and, we used TestDFSIO benchmark –comes with Hadoop– to validate our throughput results on a single server for read/write operations on HDFS.

As explained, the first input of our greedy algorithm is mutual throughput degradations ($D_{i,j}s$) that we collected by profiling data on each physical server for all combination of ten RSs (1KB-512KB) and 23 FSs (1KB-1GB) for each server; i.e., (10×23)× (10×23)=52900 individual runs in total.

TABLE III. DIFFERENT SCENARIOS CONSIDRED FOR THE PROPOSED ALGORITHM IMPLEMENT ON CLOUD PROTOTYPE

| Initial State | | | |
|---|---|---|---|
| Server1 ($M_1$) | Server2 ($M_1$) | Server3 ($M_2$) | Server4 ($M_2$) |
| (32KB, 64KB) | (32KB, 64MB) | (256KB, 1MB) | (2KB, 32KB) |
| (4KB, 16KB) | (512KB, 2MB) | (4KB, 2MB) | (512KB, 64MB) |
| (16KB, 32MB) | (128KB, 512KB) | (32KB, 8MB) | (8KB, 4MB) |
| **Sequences** | | | |
| 1 | (16KB, 64KB), (32KB, 1M), (64KB, 64MB), (32KB, 2MB), (8KB, 64MB) | | |
| 2 | (4KB, 16KB), (2KB, 16M), (2KB, 8KB), (32KB, 256KB), (16KB, 64MB) | | |
| 3 | (256KB, 2MB), (8KB, 3M), (32KB, 64MB), (4KB, 256MB), (8KB, 32MB) | | |

We also implemented the brute-force technique to gauge performance of our greedy algorithm in this work. Tables III explains our test cases in which our servers were initialized –with random combination of workloads. After that, a sequence of workloads is gradually injected into the system and allocated by our greedy algorithm. Figure 9 shows results of these experimental for three different values of α. In this figure, each bar represents the average minimum throughput of all servers. This figure shows that correct setting of α can have a great impact in minimizing the average throughput degradation of consolidated workloads in a system. For example, in this figure, α=1 and α=1.5 represent two cases in which one is too conservative (α=1), while the other is too aggressive to efficiently share LLC among concurrent workloads. The case for α=1.3, however, shows a balanced level of throughput degradation for all three sequences. We also like to highlight that, the aforementioned scenarios in Table III are not the only cases we used to evaluate the quality of our greedy algorithm. Other cases are however not reflected here as they are produced similar results to the ones we present here. In all cases, our greedy algorithm manages to find a relatively close suboptimal solution to the optimal one found by the brute-force algorithm. Also, the overhead of our developed monitoring program was always negligible compared with the CPU share of actual workloads in a system.

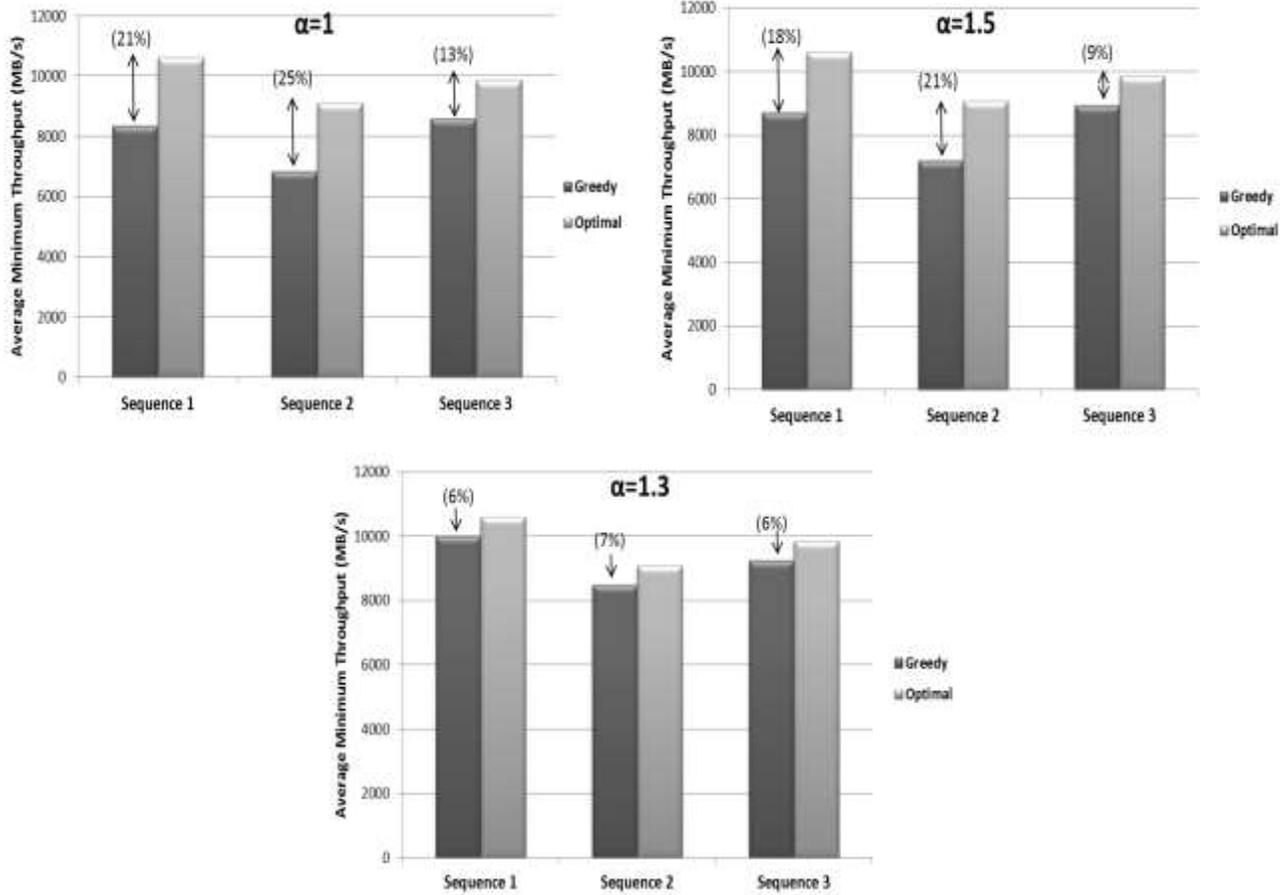

Figure 9. Comparison of optimal scheme and greedy algorithm for α=1, 1.3, and 1.5

## IX. CONCLUSION AND DISCUSSION

We investigated several challenges of efficient workload consolidation for data-intensive applications employing Hadoop distributed file system. Here, we first examined the inter-relationship between workload consolidation, resource contention, and throughput degradation on a physical server. Such examinations revealed how throughput degradation of data-intensive workloads is a function of LLC contention and mutual throughput degradation of workloads on one another. We then used the observed results and proposed two criteria to check/estimate throughput degradation of multiple workloads before consolidating them on a physical server. These criteria were then used to formulate the problem of consolidating multiple workloads on multiple servers as two-dimensional bin packing problem and also to propose our greedy approach in allocating workloads upon their arrival on systems that are already under load. Results were very promising showing that our greedy approach manages to find a relatively close suboptimal solution to the optimal one.